\documentclass{jfm}
\usepackage{graphicx}
\usepackage{epstopdf,epsfig}
\usepackage{newtxtext}
\usepackage{newtxmath}
\usepackage{natbib}
\usepackage{placeins}
\usepackage[dvipsnames]{xcolor}
\usepackage{hyperref}
\usepackage{dashrule}
\usepackage{array, multirow}
\hypersetup{
    colorlinks = true,
    urlcolor   = blue,
    citecolor  = black,
}

\newcommand{\RomanNumeralCaps}[1]

\shorttitle{On the time scales of spectral evolution of nonlinear waves}
\shortauthor{A. Simonis, A. Hrabski and Y. Pan}

\title{On the time scales of spectral evolution of nonlinear waves}

\author{Ashleigh Simonis\aff{1},
  Alexander Hrabski\aff{1}
 \and Yulin Pan\aff{1}
 \corresp{\email\href{mailto:email@address.com}{yulinpan@umich.edu}}}

\affiliation{\aff{1}Department of Naval Architecture and Marine Engineering, University of Michigan,
Ann Arbor, MI 48109, USA}

\begin{document}

\maketitle

\begin{abstract}
As presented in \citet{AnnenkovShrira09}, when a surface gravity wave field is subjected to an abrupt perturbation of external forcing, its spectrum evolves on a ``fast'' dynamic time scale of $O(\varepsilon^{-2})$, with $\varepsilon$ a measure of wave steepness. This observation poses a challenge to wave turbulence theory that predicts an evolution with a kinetic time scale of $O(\varepsilon^{-4})$. We revisit this unresolved problem by studying the same situation in the context of a one-dimensional Majda-McLaughlin-Tabak (MMT) equation with gravity wave dispersion relation. Our results show that the kinetic and dynamic time scales can both be realised, with the former and latter occurring for weaker and stronger forcing perturbations, respectively. The transition between the two regimes corresponds to a critical forcing perturbation, with which the spectral evolution time scale drops to the same order as the linear wave period (of some representative mode). Such fast spectral evolution is mainly induced by a far-from-stationary state after a sufficiently strong forcing perturbation is applied. We further develop a set-based interaction analysis to show that the inertial-range modal evolution in the studied cases is dominated by their (mostly non-local) interactions with the low-wavenumber ``condensate'' induced by the forcing perturbation. The results obtained in this work should be considered to provide significant insight into the original gravity wave problem. 

\end{abstract}


\section{Introduction}
\label{sec:introduction}
Wave turbulence theory (WTT) describes the statistical properties of ensembles of weakly nonlinear interacting waves, with rich applications in many physical contexts, e.g., ocean waves \citep{Zak67, Nazarenko16}, acoustics \citep{Lvov97}, magnetohydrodynamics \citep{Galtier2000}, quantum turbulence \citep{Nazarenko06}, and others. The centrepiece of WTT is a wave kinetic equation (WKE), which describes the time evolution of the wave action spectrum as an integral over wave-wave interactions. The WKE yields a stationary Kolmogorov-Zakharov (KZ) power-law solution associated with a constant flux, which has been observed in many wave systems.

Quantitative validations of the WKE and its predictions in different physical contexts have been a prominent topic in the wave turbulence community for decades. These studies include extensive numerical and experimental validations of the spectral slope and energy flux (or Kolmogorov constant) of the KZ solutions \citep[e.g.][]{Pan14,Hrabski2022,Zhang22a,Falcon22,Zhu23}, numerical validation of the initial spectral evolution predicted by the WKE \citep[e.g.][]{Zhu22,Banks22}, and rigorous mathematical justification of the WKE \citep[e.g.][]{Deng2021,Deng2023,Buckmaster22}. Although successes in verifying the WKE have been reported in many cases, situations where WKE predictions fail have also been identified, such as circumstances associated with finite-size effects \citep[e.g.][]{Lvov10,Hrabski2020,Zhang22b}, coherent structures in the field \citep[e.g.][]{Rumpf09}, and the strong turbulence regime \citep[e.g.][]{Chibbaro2017}.\\
\indent In spite of the significant advancement in understanding the WKE, there exist a series of relevant studies on surface gravity waves that remain largely unexplained. A representative work of this series is the paper by \citet{AnnenkovShrira09}, which considers the spectral evolution when a stationary gravity wave field is subjected to an abrupt perturbation of external wind forcing. According to the WKE of gravity waves, in the form of $\partial n/\partial t \sim n^3$ with $n\sim \varepsilon^2$ the wave action spectrum, one would expect an evolution with the kinetic time scale $O(\varepsilon^{-4})$. However, simulations of the dynamic equation, specifically the Zakharov equation \citep{Zak68}, show a faster evolution with a dynamic time scale of $O(\varepsilon^{-2})$. This ``fast'' spectral evolution is also observed in wave-tank experiments \citep[e.g.][]{Autard95, Waseda01} and field studies \citep{vanvledder93} when the wind exhibits a sudden change in speed or direction, as well as numerical simulations for the initial evolution of some wave spectra \citep{Dysthe03}. In a study by \citet{AnnenkovShrira18}, an attempt is made in explaining the fast spectral evolution using the so-called generalised WKE, but only limited success is achieved. This unexplained issue is concerning since the WKE of surface gravity waves (also known as Hasselmann's kinetic equation \citep{Hasselmann62}) is currently used in modern wave modelling codes whose reliability is pertinent to weather forecasting, climate modelling, and navigation \citep{Janssen04}.\\
\indent In this paper, we revisit the spectral evolution problem in the context of the one-dimensional (1D) Majda–McLaughlin–Tabak (MMT) equation with gravity wave dispersion relation. The MMT model is favourable in the sense that it captures the essential dynamics of wave turbulence while being exempt from the complexities associated with surface gravity waves (such as the need to remove quadratic nonlinearity terms in deriving the WKE). In addition, with a 1D model, we are able to perform large numbers of ensemble simulations, with which the statistical properties can be reliably computed through ensemble averages. We will conduct the study and analysis closely following \citet{AnnenkovShrira09}, i.e., with numerical setups and evaluation of properties as consistent as possible, except using the 1D MMT model.\\
\indent For a given stationary wave field, we show that evolution with kinetic and dynamic time scales can be observed for weaker and stronger forcing perturbations, respectively. The transition from the former to the latter regime corresponds to a forcing perturbation that triggers a spectral evolution time scale comparable to the linear time scale, i.e., the wave period of some representative mode, violating the basis of the WKE. We further show, through a study varying the energy of the base stationary wave field, that this violation of time scale separation is primarily a result of the spectrum far from the stationary state after a sufficiently strong forcing perturbation is applied, rather than the increase of overall nonlinearity level of the wave field. We finally develop a set-based interaction analysis, which allows us to understand that the spectral evolution for modes in the inertial range is predominantly governed by their (mostly non-local) interactions with the low-wavenumber condensate (or regions sufficiently filled with energy) induced by the forcing perturbation.    

\section{Methodology}\label{methodology}

\subsection{The MMT model}
\label{sec:mmt_model}
We study the evolution of random wave fields through the one-dimensional MMT equation \citep{MMT1997}, which is a family of nonlinear dispersive wave equations widely used to study wave turbulence problems \citep[e.g.][]{Cai1999,Zak04,Chibbaro2017,Hrabski2022},

\begin{equation}
  \textnormal{i}\frac{\partial\psi}{\partial t}={\lvert{\partial _x}\rvert}^\alpha \psi+\lambda{\lvert{\partial _x}\rvert}^{\beta/4}({\big\lvert{\lvert{\partial _x}\rvert}^{\beta/4}\psi\big\rvert}^2{\lvert{\partial _x}\rvert}^{\beta/4}\psi),
  \label{MMT}
\end{equation}
where $\psi(x,t)$ is a field taking complex values. The parameter $\beta$ controls the nonlinearity formulation and $\alpha$ controls the dispersion relation $\omega(k)=|k|^\alpha$ with $\omega$ the frequency and $k$ the wavenumber. $\lambda=-1$ and $1$ represent the focusing (defocusing) nonlinearity, respectively \citep{Cai2001}. The MMT equation \eqref{MMT} conserves both the total Hamiltonian and wave action. In our study, we use $\alpha=1/2$ to mimic the dispersion of surface gravity waves, and use $\beta=0$ for convenience which is consistent with a portion of the original study of the MMT equation \citep{MMT1997} and is widely used in other studies \citep[e.g.][]{Cai1999,Rumpf13,Rumpf15}. We note that although $\beta=0$ is not entirely representative of surface gravity waves, the primary conclusion of the paper does not depend on a specific choice of $\beta$. We will present the results for the defocusing case ($\lambda=1$) in the main paper, and those for the focusing case in the Appendix \ref{AppB}, with both cases exhibiting similar physics regarding the spectral evolution time scales.

From a standard wave turbulence consideration, a statistical description of the wave field can be obtained by defining the wave action spectrum $n_{k}=\langle{\lvert{\hat{\psi}_k}\rvert}^2\rangle$, with $\hat{\psi}_k$ the Fourier transform of $\psi$ and the angle brackets denoting an ensemble average. Under conditions of weak nonlinearity, random phases, and infinite domain, the time evolution of $n_k$ is governed by the wave kinetic equation (WKE) (for $\beta=0$),
\begin{equation}
\begin{split}
  \frac{\partial n_{k}}{\partial t}= 4\pi\int (n_{1}n_{2}n_{3}+n_{1}n_{2}n_{k}-n_{1}n_{3}n_{k}-n_{2}n_{3}n_{k})
  \\
  \times\delta(\omega_{1}+\omega_{2}-\omega_{3}-\omega)\delta(k_{1}+k_{2}-k_{3}-k)\textnormal{d}k_{1}\textnormal{d}k_{2}\textnormal{d}k_{3}
  \label{WKE}
\end{split}
\end{equation}
where $\delta$ is the Dirac delta function. According to \eqref{WKE}, the spectrum evolves with a kinetic time scale, i.e., $n_k$ experiences significant change with a time scale $O(\varepsilon^{-4})$, and $\partial n_k/\partial t\sim O(\varepsilon ^6)$ over the evolution, with $\varepsilon \sim \sqrt{n_k}$ a measure of wave steepness. Hereafter, we will refer to these relations as kinetic scaling in this paper, which is in contrast to the dynamic scaling of a time scale $O(\varepsilon^{-2})$ and $\partial n_k/\partial t \sim O(\varepsilon ^4)$ that one can directly obtain from \eqref{MMT}.    
\subsection{Numerical procedure}
\label{sec:num_proc}
We simulate \eqref{MMT} with 4096 modes (before de-aliasing) on a periodic domain of size $L=2\pi$, with the addition of forcing and dissipation terms. The forcing is in white-noise form, given by
\begin{equation}
  F =
    \begin{cases}
      F_{r}+\textnormal{i}F_{i}, & 4\le k \le 13,\\
      0, & \text{otherwise},
    \end{cases}
    \label{F}
\end{equation}
with $F_r$ and $F_i$ independently drawn from a Gaussian distribution $\mathcal{N}(0, \sigma^2)$. The dissipation is introduced with the addition of two hyperviscosity terms
\begin{equation}
  \left.\begin{aligned}
  D_{1} =
    \begin{cases}
      -\textnormal{i}\nu_{1}\hat{\psi}_{k}, & k\ge 900,\\
      0, & \text{otherwise},
    \end{cases} \\
    D_{2} =
    \begin{cases}
      -\textnormal{i}\nu_{2}\hat{\psi}_{k}, & k\le 4,\\
      0, & \text{otherwise},
    \end{cases} 
    \label{D}
\end{aligned}\right\}
\end{equation}
at small and large scales, respectively. Since the MMT model supports the inverse cascade, it is necessary to use large-scale dissipation to avoid energy accumulation at large scales. The dissipation coefficients are fixed to be $\nu_{1}=1\times 10^{-14}(k-900)^8$ and $\nu_{2}=3k^{-4}$ for all numerical experiments. The numerical schemes used for the simulation are discussed in detail in previous papers \citep{Hrabski2020,Hrabski2022}.

In order to reproduce the physical scenario as in \citet{AnnenkovShrira09}, we perform the simulation with two stages. In stage 1, we simulate \eqref{MMT} with \eqref{F} and \eqref{D} using $\sigma=\sigma_0$ for sufficient time to reach a stationary wave field (and spectrum), starting from an initial spectrum that exponentially decays in $|k|$. In stage 2, we add a perturbation $\Delta \sigma$ to the forcing magnitude (leading to $\sigma=\sigma_0+\Delta \sigma$) and study how the spectrum evolves subject to the perturbation. In both stages, we use a time step $\Delta t=0.012$. For the main cases studied in this paper, we consider the stationary state obtained with forcing $\sigma_0=0.004$ in stage 1, and consider forcing perturbations $\Delta \sigma \in [0.033,1.461]$ in stage 2 that is sufficient to cover the physical regimes of our interest. The process is simulated with an ensemble of 65000 simulations with different random seeds in the forcing, and all statistical results related to $n_k$ presented below are obtained directly from the average over the whole or part of the ensemble that is statistically convergent. 

\section{Results}
\label{sec:results}
\begin{figure}
  \centerline{\includegraphics{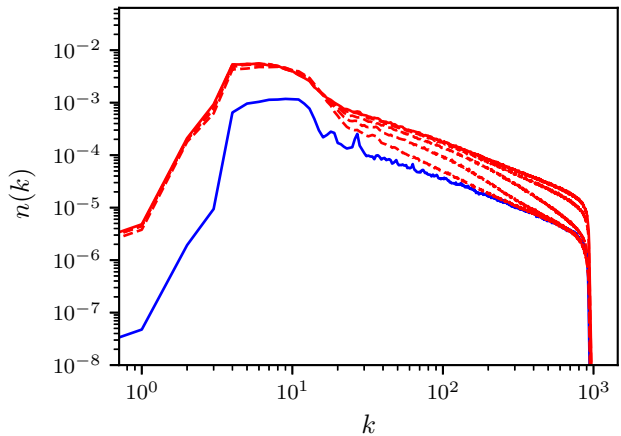}}
  \caption{A typical case of the evolution of wave action spectrum, where $n(k,0)$ ({\color{blue}{---}}) and $n(k,t_f)$ (\begingroup\color{red}{---}\endgroup) represent the stationary spectra before the forcing perturbation (stage 1) and after sufficient evolution with forcing perturbation applied (stage 2). The intermediate spectra in the evolution between $n(k,0)$ and $n(k,t_f)$ are plotted (\begingroup\color{red}{$---$}\endgroup).}
\label{fig:fig1}
\end{figure}

Figure \ref{fig:fig1} shows the evolution of a spectrum in a typical case with $\sigma_0=0.004$ and $\Delta\sigma = 0.033$. We see that a stationary power-law spectrum forms at the end of stage 1, which serves as the base state of the problem. As a forcing perturbation is excited in stage 2, the forcing scales (as well as scales slightly larger than that) first experience an abrupt growth, forming a condensate at large scales. The growth is then propagated to smaller scales, which eventually fill in the full spectral range and form the final stationary power-law spectrum. We note that such behaviour regarding the propagation of growth is closely related to the fact that \eqref{MMT} under selected values of $\alpha$ and $\beta$ is an infinite capacity system \citep{Nazarenko11,Newell11}. Hereafter, for simplicity, we assign time $t=0$ to the stationary state before a forcing perturbation is applied, and $t=t_f$ to the time when the final stationary state is formed. We measure the wave steepness of each state as $\varepsilon=\sqrt{E_L}$ with $E_L$ the linear energy (or Hamiltonian) of the system. Accordingly, we use $\varepsilon_0$ and $\varepsilon_f$ as the wave steepness at $t=0$ and $t=t_f$. We note that this definition of $\varepsilon$ is consistent with that in \citet{AnnenkovShrira09} with the additional factor of constant peak wavenumber $k_p$ omitted in our definition. 

\begin{figure}
  \centerline{\includegraphics{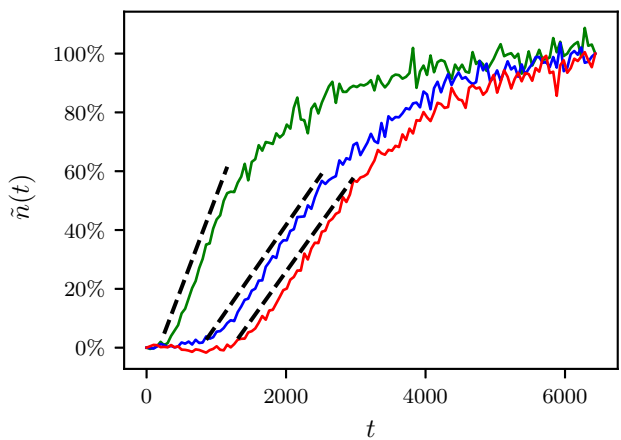}}
  \caption{A typical case of the evolution of modal wave action at three wavenumbers $k=150$ (\begingroup\color{ForestGreen}{---}\endgroup), $k=500$ (\begingroup\color{blue}{---}\endgroup), $k=800$ (\begingroup\color{red}{---}\endgroup) in the inertial-range. The growth rates $\partial \tilde{n}_k/\partial t$ measured over the range of $\tilde{n}_k \in [5\%, 40\%]$ are indicated (\begingroup{$---$}\endgroup).}
\label{fig:fig2}
\end{figure}

We are interested in the growth rate in the transient period in stage 2, based on which we can evaluate $\partial n_k/\partial t$ and the associated scaling with wave steepness. For this purpose, we follow \citet{AnnenkovShrira09} to measure $\partial n_k/\partial t$ for a given mode $k$ using data over the interval of 5-40\% of the total modal growth. More specifically, if we define a normalised wave action spectrum $\tilde{n}(k,t)=(n(k,t)-n(k,0))/(n(k,t_f)-n(k,0))$, we can then measure $\partial \tilde{n}_k/\partial t$ over the range of $\tilde{n}_k \in [5\%, 40\%]$ via a least-square fit, and then scale back to compute $\partial n_k/\partial t$. Figure \ref{fig:fig2} shows the evolution of $\tilde{n}(k,t)$ for three selected modes in the inertial range, as well as the growth rate evaluated using the $[5\%, 40\%]$ interval. It is clear that the evaluated growth rate over the chosen interval is sufficient to capture the instantaneous spectral growth after the forcing perturbation is imposed. 

\subsection{Scaling of spectral growth rate with $\varepsilon$}
\label{sec:sec3-1}
\begin{figure}
  \centerline{\includegraphics{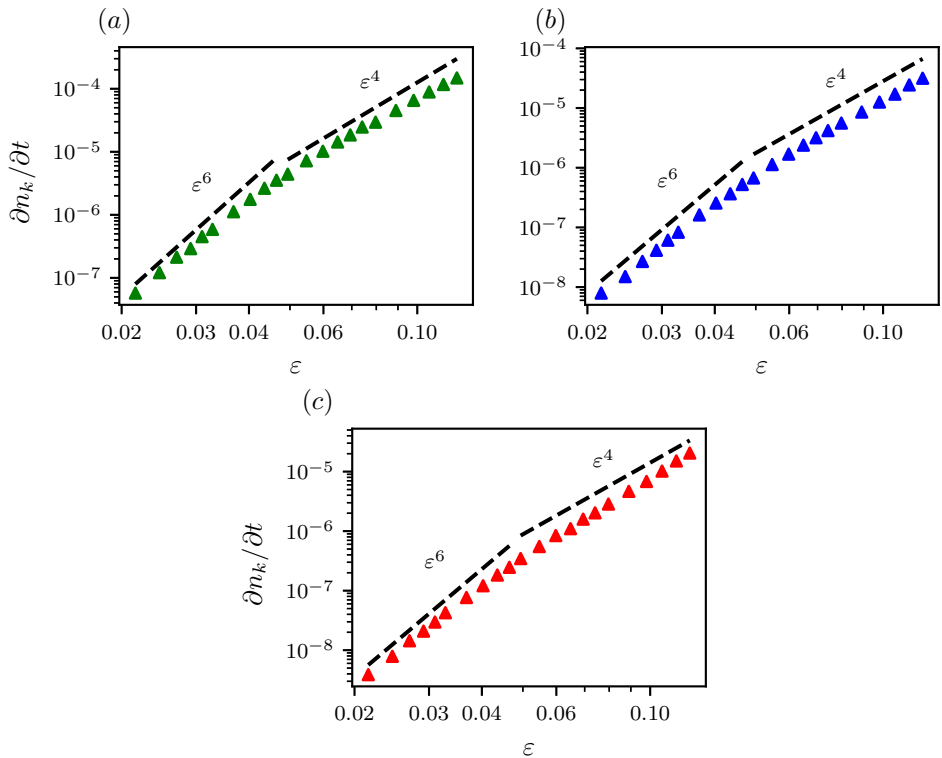}}
  \caption{Rate of spectral evolution $\partial n_k/\partial t$ as a function of nonlinearity $\varepsilon$ for ({\it a}) $k=150$ (\begingroup\color{ForestGreen}$\blacktriangle$\endgroup), ({\it b}) $k=500$ (\begingroup\color{blue}$\blacktriangle$\endgroup),  and ({\it c}) $k=800$ (\begingroup\color{red}$\blacktriangle$\endgroup), with the kinetic $\varepsilon^{6}$ and dynamic $\varepsilon^{4}$ scalings indicated by the dashed lines.}
\label{fig:fig3}
\end{figure}

To evaluate the scaling of $\partial n_k/\partial t$ with the wave steepness $\varepsilon$, we use the final-state $\varepsilon=\varepsilon_f$ as a representative value of the wave steepness for the case \citep[cf.][]{AnnenkovShrira09}. Figure \ref{fig:fig3} shows $\partial n_k/\partial t$ of three modes in the inertial range for a broad range of $\varepsilon \in [0.022,0.124]$, obtained from 22 cases with $\Delta \sigma \in [0.033,1.461]$ starting from a base state with $\varepsilon_0=0.010$ resulting from $\sigma_0=0.004$. 
As we see in figure \ref{fig:fig3}, the kinetic scaling $\partial n_k/\partial t \sim O(\varepsilon^{6})$ is realised in the range of $\varepsilon \lesssim 0.05$, corresponding to small $\Delta \sigma$ with $\Delta \sigma \lesssim 0.36$. For larger forcing perturbations, we find dynamic scaling in the range of $\varepsilon \gtrsim 0.05$. We remark that in the previous work for gravity waves (Annenkov \& Shrira \citeyear{AnnenkovShrira09}), only dynamic scaling is observed. This is possibly due to their studies only being conducted for strong forcing perturbations, or other reasons that we will leave for future study.   

We next investigate why the ``fast'' dynamic scaling becomes relevant for the spectral evolution, a critical question that is not answered in previous works \citep[e.g.][]{AnnenkovShrira06,AnnenkovShrira09,AnnenkovShrira18}. The realisation of dynamic scaling indicates that the WKE \eqref{WKE} (or more generally, wave turbulence theory) must break down. One situation in which this could happen is a violation of time scale separation, i.e., when the nonlinear modal time scale becomes comparable to the linear modal time scale. In particular, the linear time scale of a mode is given by the modal wave period
\begin{equation}
    \tau_{L}=\frac{2\pi}{\omega}.
    \label{tau_l}
\end{equation}

The nonlinear time scale can be computed in our case as the spectral evolution time scale
\begin{equation}
    \tau_{NL}=\frac{n_{k}}{\frac{\partial n_{k}}{\partial t}},
    \label{tau_nl}
\end{equation}
with $\partial n_k/\partial t$ evaluated from our numerical data. Hereafter, we will also use the terminology ``spectral evolution time scale'' which is somewhat more illuminating than ``nonlinear time scale''. In addition, we note that in many cases \citep[e.g.][]{Newell01, Newell11} the nonlinear time scale is taken as the kinetic time scale, which is not appropriate here since $\partial n_k/\partial t$ in \eqref{tau_nl} is evaluated by the dynamic equation instead of the WKE. The ratio of the spectral evolution and linear time scales is given by 
\begin{equation}
    \rho=\frac{\tau_{NL}}{\tau_{L}}.
    \label{ts_ratio}
\end{equation}
The WKE is expected to be valid only when the time scales are well separated with $\rho \gg O(1)$. 

\begin{figure}
  \centerline{\includegraphics{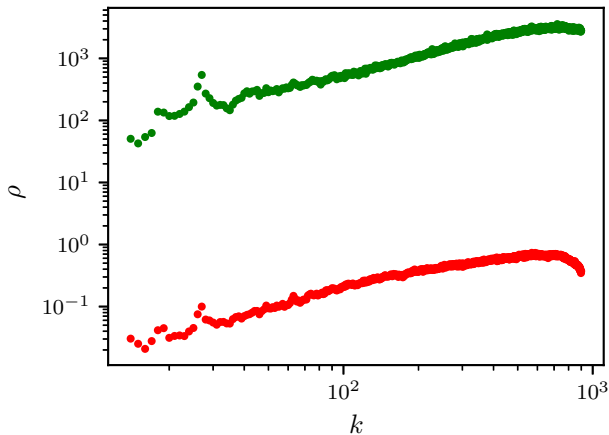}}
  \caption{Ratio of time scales $\rho$ as a function of $k$ for cases with $\Delta\sigma=0.033$, $\varepsilon=0.022$ (\begingroup\color{ForestGreen}$\bullet$\endgroup) and $\Delta\sigma=1.461$, $\varepsilon=0.124$ (\begingroup\color{red}$\bullet$\endgroup).} 
\label{fig:fig4}
\end{figure}

\begin{figure}
  \centerline{\includegraphics{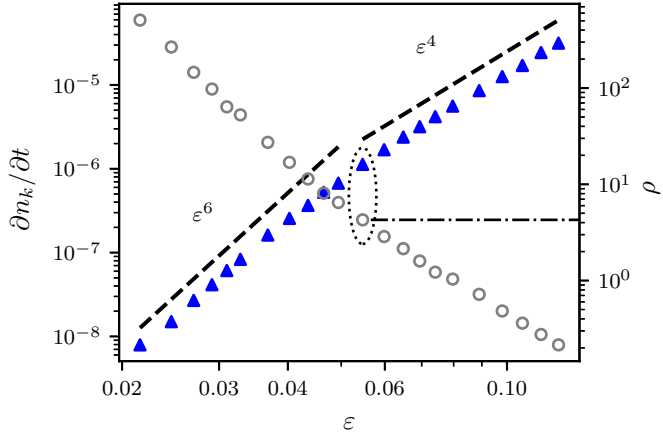}}
  \caption{Ratio of the time scales $\partial n_k/\partial t(k=500)$ (left axis, \begingroup\color{blue}$\blacktriangle$\endgroup) and $\rho(k=100)$ (right axis, \begingroup\color{gray}$\circ$\endgroup) as functions of $\varepsilon$. The region of transition from kinetic to dynamic scaling is circled, with the corresponding value of $\rho$ indicated (\begingroup{$-\cdot-$}\endgroup).}
\label{fig:fig5}
\end{figure}

To study the time scale separation in our cases, we first mention that $\rho$ is, in general, a function of $k$, as shown in two typical cases in figure \ref{fig:fig4} for $\Delta \sigma=0.033$ and $1.461$. We see that $\rho$ generally increases with $k$ in a power-law form, which is related to our choice $\beta=0$ in \eqref{MMT} (which makes the nonlinear strength weaker for high $k$ than that from a positive $\beta$). Additionally, we see that as $\Delta \sigma$ increases from the former values to the latter, $\rho$ generally drops below $O(1)$, indicating the breakdown of the WKE in the latter case. We next seek to understand the relationship between the time scale separation and the transition to the dynamic scaling of spectral evolution. To this end, we first note that the transition to dynamic scaling at all inertial-range wavenumbers (see figure \ref{fig:fig3}) occurs at a similar forcing perturbation, indicating that the WKE breaks down for the overall spectral range instead of a particular wavenumber. Therefore, we evaluate $\rho$ at a representative wavenumber $k=100$, which leads to a relatively low value of $\rho$ in the inertial range as an assessment of the overall validity of WTT. We plot in figure \ref{fig:fig5} $\rho(k=100)$ as a function of $\varepsilon$, overlaid with the previous plot of $\partial n_{k}/\partial t$ for $k=500$ (as an example). We see that the transition to dynamic scaling occurs at $\rho \approx 4 \sim O(1)$, which indicates that the dynamic scaling range is consistent with the failure of the WKE due to the violation of time scale separation. We also remark that if a different value of $k$ is chosen for the evaluation of $\rho$, we will end up with a slightly different transition value of $\rho$ which is at most approximately 10 as one can estimate from figure \ref{fig:fig4}.

\begin{figure}
  \centerline{\includegraphics{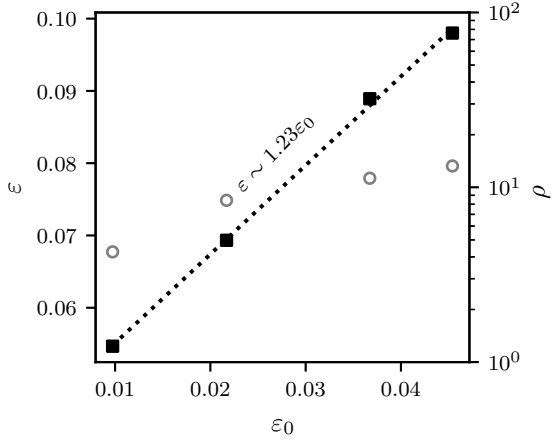}}
  \caption{Values of $\varepsilon$ (left axis, $\blacksquare$) and $\rho(k=100)$ (right axis, \begingroup\color{gray}$\circ$\endgroup) corresponding to transition to dynamic scaling for each base nonlinearity level $\varepsilon_{0}$, with a linear fitting ($\cdots$) and relation indicated.}
\label{fig:fig6}
\end{figure}

While the above analysis reveals the reduction of $\tau_{NL}$ as the underlying reason for the transition to dynamic scaling, the cause of this reduction remains unclear. In many previous studies, the reduction of $\tau_{NL}$ is attributed to the increase of nonlinearity level, which transits the dynamics into a strong turbulence regime (often associated with intermittency) where the WKE becomes irrelevant. However, we argue that this is not the major reason for the transition to dynamic scaling observed in our study, which is instead mainly triggered by the spectrum being too far from the stationary spectrum after a strong forcing perturbation is applied. To provide evidence for this argument, we repeat our analysis for three additional situations with higher values of $\varepsilon_0$. Figure \ref{fig:fig6} shows that the transition value of $\varepsilon$ increases with $\varepsilon_0$ following an approximately linear form of $\varepsilon \sim \varepsilon_0$, and that these transitions all correspond to $\rho(k=100) \sim O(1-10)$, which is in agreement with the above analysis. This is consistent with the argument about the transition caused by a far-from-stationary spectrum induced by a strong forcing perturbation, since the deviation of the spectrum from the stationary state is approximately measured by the difference between $\varepsilon$ and $\varepsilon_0$, which does not change significantly in all cases. Moreover, figure \ref{fig:fig6} is in clear contradiction with a transition induced by a high nonlinearity level (or strong turbulence), in which case one would otherwise expect that the transition occurs at approximately the same $\varepsilon$ for different $\varepsilon_0$.  

The failure of the WKE due to the spectrum far from the stationary state can also be reasoned from a thought experiment: given a spectrum, one can assume \emph{a priori} that the WKE is valid, based on which $\partial n_{k}/\partial t$ can be computed. Such computed $\partial n_{k}/\partial t$ may \emph{a posteriori} be found to violate the condition of $\rho \gg O(1)$ that invalidates the assumed WKE (see examples for internal gravity waves in \citet{Lvov12} and \citet{Eden19}). In this view, as the spectrum deviates from the stationary state, there exists a critical spectral form (even if the nonlinearity level is relatively low) that drives fast evolution beyond which the WKE fails.

\subsection{Dominant interactions in modal growth}
From section {\S} \ref{sec:sec3-1} we understand that the forcing perturbation (and the associated condensate) creates a deviation of the spectrum from the stationary state, which drives the subsequent spectral evolution. It is therefore reasonable to further argue that the inertial-range modal growth is dominated by direct interaction with the condensate peak, or at least non-local interactions with large-scale features. In this section, we verify this hypothesis using a new set-based interaction analysis. We remark that this is not a trivial task in the sense that these interactions cannot be analysed on the basis of the WKE, but rather must be analysed based on the dynamic equation \eqref{MMT} which is valid for all cases. 

To start, we first define a set-based modal growth rate $\partial n_k/\partial t\rvert_{A}$ to be the evolution rate of $n_k$ due to the interaction of mode $k$ with $k_i\in A$ for $i=1,2,3$. By definition, if $A=\Lambda \equiv [-k_{max}, k_{max}]$ (i.e., the full spectral range with $k_{max}=1024$ after de-aliasing), then $\partial n(k)/\partial t\rvert_{A}=\partial n(k)/\partial t$. For $A\in \Lambda$, we can compute $\partial n(k)/\partial t\rvert_{A}$ by invoking \eqref{MMT} as
\begin{equation}
    \frac{\partial n(k)}{\partial t}\biggr\rvert_{A} = \sum_{k_{1} \in A, k_{2}\in A, k_{3}\in A} 2\textnormal{Im} \langle\hat{\psi}_{1}\hat{\psi}_{2}\hat{\psi}^\ast_{3}\hat{\psi}^\ast_{k}\rangle\delta^{12}_{3k},
    \label{dnk_a}
\end{equation}
where $\delta^{12}_{3k}\equiv \delta(k_1+k_2-k-k_3)$ and the ${\hat{\psi}}_k^*$ denotes the complex conjugate of $\psi_k$. The direct computation of \eqref{dnk_a} in spectral space is very expensive, with a computational complexity of $O(k_{max}^3)$ for each ensemble member of the group at each time instant (the ensemble and the time average are then used to compute the operator $\langle \, \rangle$). The computational cost can be significantly reduced if the evaluation of \eqref{dnk_a} can be performed in physical space, which turns out to be possible as we detail in Appendix \ref{appA}, with the result
\begin{equation}
    \frac{\partial n(k)}{\partial t}\biggr\rvert_{A}\equiv 2\textnormal{Im} \langle \hat{\psi}^*_{k}    
    \mathcal{F} [B_{A}(\psi(x))B_{A}(\psi(x))B_{A}(\psi^*(x))] \rangle,
    \label{dnk_final}
\end{equation}
where $\mathcal{F}[\,]$ represents the Fourier transform, and $B_A$ represents a Fourier-domain filter to select modes in $A$ for a quantity in physical domain. The computation of \eqref{dnk_final} only requires a number of fast Fourier transforms which is much less expensive than that for \eqref{dnk_a}.

\begin{figure}
    \centering
    \includegraphics{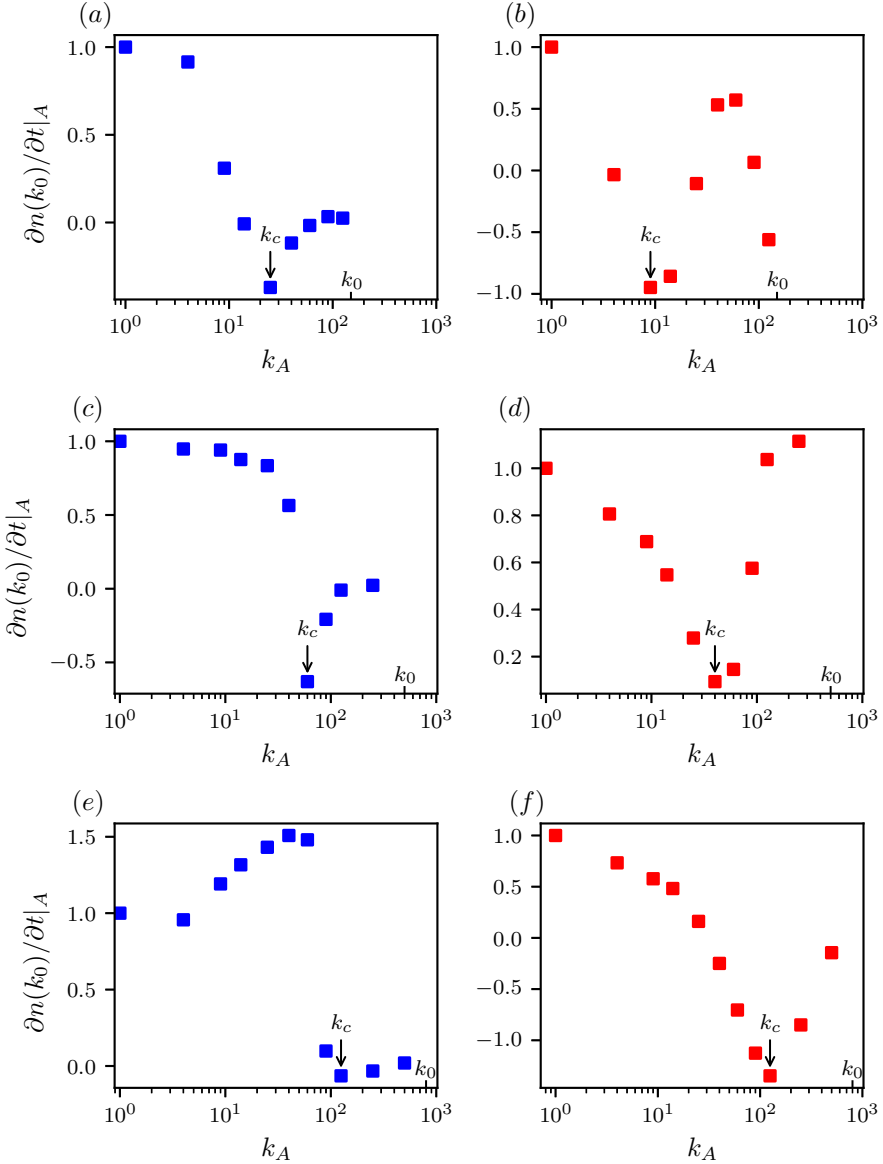}
    \caption{Rate of spectral evolution $\partial n(k_0)/\partial t\rvert_{A}$, normalized by $\partial n(k_0)/\partial t$, as a function of $k_A$, for three wavenumbers ({\it a}, {\it b}) $k_0=150$, ({\it c}, {\it d}) $k_0=500$ and ({\it e}, {\it f}) $k_0=800$. The columns on the left ({\it a}, {\it c}, {\it e}) and on the right ({\it b}, {\it d}, {\it f}) are for cases with $\Delta \sigma=0.033$ and $1.461$ in the kinetic and dynamic scaling regimes, respectively. The positions of $k_c$ (where minimum value $\partial n(k_0)/\partial t\rvert_{A}$ is achieved) and $k_0$ are indicated in each sub-figure.}
    \label{fig:fig7}
\end{figure}

In order to sort out the dominant interactions that lead to the growth of a mode $k_0$, we evaluate $\partial n(k_0)/\partial t\rvert_{A}$ by setting $A=[-k_{max}, -k_A]\cup[k_A, k_{max}]$ with $k_A$ varying in $[1,k_0]$. Therefore, for fixed $k_0$, $\partial n(k_0)/\partial t\rvert_{A}$ can be considered as a function of $k_A$. As $k_A$ decreases from $k_0$ to 1, $\partial n(k_0)/\partial t\rvert_{A}$ accounts for more non-local interactions with large scales (which is more important than interactions with small scales that are not particularly studied in current setting). Figure \ref{fig:fig7} plots $\partial n(k_0)/\partial t\rvert_{A}$ (normalised by $\partial n(k_0)/\partial t$) as a function of $k_A$ for three selected values of $k_0$ in the inertial range, and for two cases of $\varepsilon=0.022$ and $0.124$ corresponding to regimes of kinetic and dynamic scaling, respectively. We first see that, for each case, the leftmost point (i.e., when $k_A=1$) recovers the total $\partial n(k_0)/\partial t$, which can be considered as a validation of our computation. For $k_A$ relatively close to $k_0$, local interactions are represented, with those for higher nonlinearity levels showing stronger fluctuations, i.e., stronger local interactions. Nevertheless, for all cases, there exists a wavenumber $k_c$ (marked in each figure) at which the normalised $\partial n(k_0)/\partial t\rvert_{A}$ becomes close to zero or negative. This means that the overall local interactions in the range $[k_c, k_0]$ do not contribute much (or even contribute negatively) to the total $\partial n(k_0)/\partial t$. The major contribution to recover $\partial n(k_0)/\partial t$ occurs for a range immediately left of $k_c$, where the normalised $\partial n(k_0)/\partial t\rvert_{A}$ grows quickly from the minimum value to 1. This range features non-local interactions with scale separation of $k_c/k_0\sim O(0.1)$ (see table \ref{tab:table1} for exact numbers) for all cases shown in the figure. The fact that $k_c$ increases with $k_0$ indicates that the large scales participating in the dominant interactions propagate to higher wavenumbers. For $k_0=150$, the non-local interactions mainly involve the forced condensate range up to about $k_c\sim O(10)$. For $k_0=500$ and $k_0=800$, the small-wavenumber range sufficiently filled with energy during the propagation of the perturbation becomes the dominant interacting modes. 
\begin{table}
\centering
    \begin{tabular}{c c c c}
    Nonlinearity $\varepsilon$ & Wavenumber of interest $k_0$ & Wavenumber at minimum $k_{c}$ & $k_{c}/k_0$   \\

    \multirow{3}{*}{0.022} & 150 & 25 & 0.167 \\
                            & 500 & 60 & 0.120 \\
                            & 800 & 125 & 0.156 \\
    \multirow{3}{*}{0.124} & 150 & 9 & 0.060 \\
                            & 500 & 40 & 0.080 \\
                            & 800 & 125 & 0.156

    \end{tabular}
\caption{\label{tab:table1} Values of $k_c$ and $k_c/k_0$ for all test cases.}
\end{table}

In summary, the analysis here implies that, for all forcing perturbations considered, the inertial-range modal growth is mainly driven by the non-local interactions of the modes with large scales. The local interactions introduce some fluctuations that are stronger for higher-nonlinearity cases, but are sufficiently decayed when enlarging the spectral range around $k_0$ and before the dominant non-local interactions take place. We also mention that in the low-nonlinearity cases, the non-local interactions (especially those with the condensate) can be approximated by a diffusion equation recently derived in \citet{Korotkevich23} for four-wave systems. 

\section{Conclusion and discussions}

In this paper, we numerically study the time scales of spectral evolution of nonlinear waves under perturbed forcing, in order to understand the previously identified ``fast'' spectral evolution with the dynamic time scale contradicting the prediction from the WKE. Our study is conducted in the context of 1D MMT equation, that mimics the wave turbulence of surface gravity waves free of other associated complexities. We show that both kinetic and dynamic time scales can become relevant for spectral evolution depending on the magnitude of the perturbed forcing applied to the stationary spectrum. The kinetic scaling occurs at weaker forcing perturbations, which transits to dynamic scaling as the forcing perturbation becomes strong enough, with the transition corresponding to the situation of the spectral evolution (or nonlinear) time scale dropping to the same level as the linear time scale. Such decrease in nonlinear time scale is mainly induced by the spectrum being far from the stationary state after the forcing perturbation is applied, instead of solely due to the usually-considered cause of increasing nonlinearity level. Finally, through a new set-based interaction analysis, we find that the inertial-range spectral growth is dominated by its nonlocal interactions with the forced condensate or large scales sufficiently filled with energy throughout all nonlinearity levels.

Our study suggests that in assessing the validity of the WKE for transient spectra, the specific spectral forms need to be considered, in conjunction with the traditionally considered factors such as weak/strong turbulence and intermittency. Although we will consider another study directly regarding surface gravity waves as our future work, the study here itself provides significant insight into the observed dynamic time scale in the spectral evolution of surface gravity waves under perturbed forcing \citep{AnnenkovShrira09}. The established interpretation may also apply to the initial evolution of the wave spectrum, which is sometimes reported to be on the dynamic scale \citep{Dysthe03} and sometimes precisely matches the prediction by the WKE \citep{Zhu22}. 

\section{Citations and references}


\textbf {Acknowledgements.} {The authors acknowledge the Simons Foundation for funding support for this work, and thank Prof. Miguel Onorato for helpful discussions on this work during the Simons Collaboration on Wave Turbulence Annual Meeting in 2022.}


\textbf {Declaration of interests.} {The authors report no conflict of interest.}




\appendix

\section{Fast computation for set-based interaction analysis}\label{appA}
We start by re-ordering \eqref{dnk_a} (by interchanging allowed operations) as
\begin{equation}
    \frac{\partial n(k)}{\partial t}\biggr\rvert_{A} = 2\textnormal{Im} \langle \hat{\psi}^*_{k}    
    \sum_{k_{1} \in A, k_{2}\in A, k_{3}\in A}  \hat{\psi}_{1}\hat{\psi}_{2}\hat{\psi}^*_{3} \delta^{12}_{3k} \rangle.
    \label{dnk_a_2}
\end{equation}
The significant computational cost comes from the summation term on the right-hand side of \eqref{dnk_a_2}. To reduce the computational cost, we can consider the summation term as the Fourier transform of a physical-space quantity:
\begin{equation}
   \sum_{k_{1} \in A, k_{2}\in A, k_{3}\in A}  \hat{\psi}_{1}\hat{\psi}_{2}\hat{\psi}^*_{3}\delta^{12}_{3k} = \mathcal{F} [B_{A}(\psi(x))B_{A}(\psi(x))B_{A}(\psi^*(x))],
   \label{pf}
\end{equation}
where $\mathcal{F}[\,]$ represents the Fourier transform, and $B_A$ represents a Fourier-domain filter to select modes in $A$ for a physical domain function $\psi(x)$. Equation \eqref{pf} can be understood from the convolution theorem for three functions, i.e., spectral domain convolution is equal to the physical domain multiplication. One can also directly confirm that this is true by starting from the right-hand side and expressing each $B_{A}(\psi(x))$ as the summation of Fourier modes:
\begin{equation}
   \frac{1}{2\pi}\int e^{-ikx} \sum_{k_1\in A} \hat{\psi}(k_1) e^{ik_1x} \sum_{k_2\in A} \hat{\psi}(k_2) e^{ik_2x} \sum_{k_3\in A} \hat{\psi}^*(k_3) e^{-ik_3x} dx
\end{equation}
which can be re-organized as
\begin{equation}
   \frac{1}{2\pi}\int \sum_{k_{1} \in A, k_{2}\in A, k_{3}\in A} \hat{\psi}(k_1) \hat{\psi}(k_2) \hat{\psi}^*(k_3) e^{i(k_1+k_2-k-k_3)x} dx.
   \label{deriv}
\end{equation}
We see that \eqref{deriv} is the same as the left hand side of \eqref{dnk_a_2} after considering $\frac{1}{2\pi}\int  e^{i(k_1+k_2-k-k_3)x} = \delta(k_1+k_2-k-k_3)$.

\section{Results from cases with focusing nonlinearity}\label{AppB}

\begin{figure}
  \centerline{\includegraphics{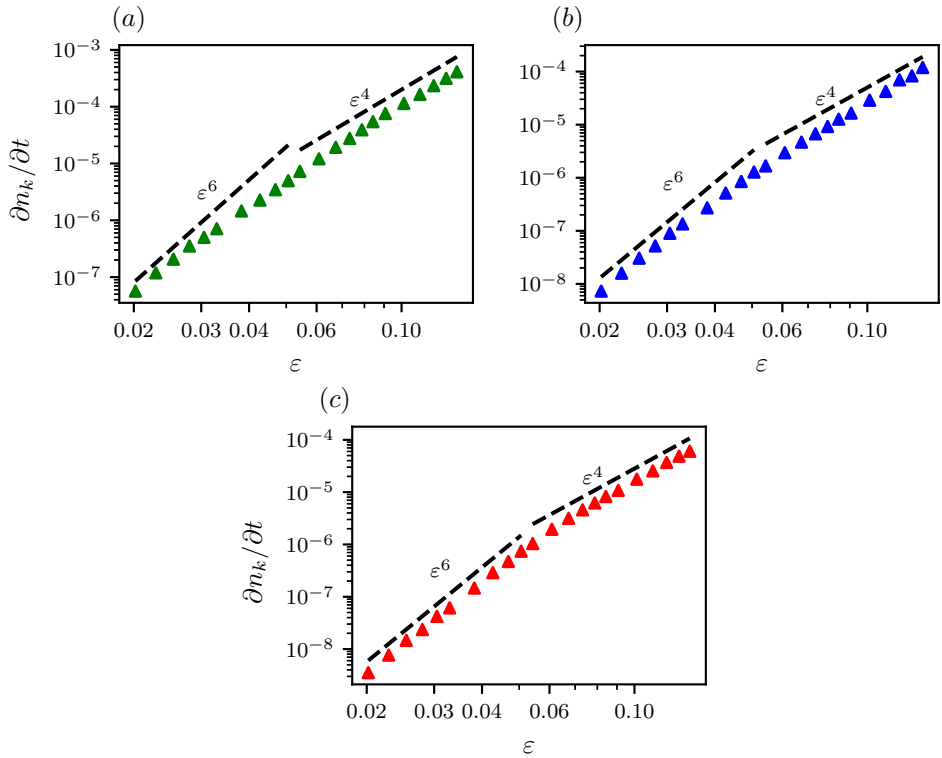}}
  \caption{See caption for figure \ref{fig:fig3}, but for the focusing case.}
\label{fig:fig8}
\end{figure}

\begin{figure}
  \centerline{\includegraphics{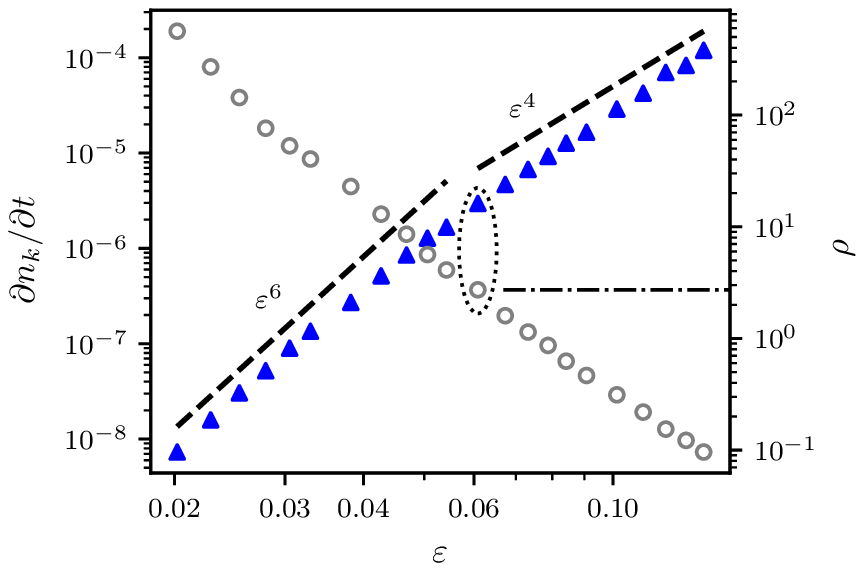}}
  \caption{See caption for figure \ref{fig:fig4}, but for the focusing case.}
\label{fig:fig9}
\end{figure}

In this appendix, we summarize results from cases with focusing nonlinearity, i.e., $\lambda=-1$ in \eqref{MMT}. All other parameters in the study are kept consistent with those in the defocusing cases reported in the main paper. Figures \ref{fig:fig8} and \ref{fig:fig9} show respectively the scaling of $\partial n_k/\partial t$ with $\varepsilon$ and the study regarding $\rho$, as counterparts of figure \ref{fig:fig3} and \ref{fig:fig4} in the main paper. It is clear from these figures that the main conclusion made for the defocusing case also applies to the focusing case. 

\FloatBarrier
\bibliographystyle{jfm}


\end{document}